\documentclass[superscriptaddress, onecolumn, amsmath, amssymb, 11pt, aps]{revtex4-2}

\usepackage{graphicx}
\usepackage{dcolumn}
\usepackage{float}
\usepackage{bm}
\usepackage{amsmath}
\usepackage[colorlinks,linkcolor=blue,anchorcolor=blue,citecolor=blue]{hyperref}
\usepackage{color}
\usepackage[markup=underlined]{changes}

\begin{document}

\title{Real-space Hybrid Topological Singularities in Structured Elastic Waves}

\author{Tong Fu}
\thanks{These authors contributed equally to this work.}
\affiliation{Department of Physics, City University of Hong Kong, Tat Chee Avenue, Kowloon, Hong Kong, China}

\author{Pengfei Zhao}
\thanks{These authors contributed equally to this work.}
\affiliation{Department of Physics, The Hong Kong University of Science and Technology, Kowloon, Hong Kong, China}

\author{Liyou Luo}
\affiliation{Department of Physics, The Hong Kong University of Science and Technology, Kowloon, Hong Kong, China}

\author{Zhiling Zhou}
\affiliation{Department of Physics, City University of Hong Kong, Tat Chee Avenue, Kowloon, Hong Kong, China}

\author{Dong Liu}
\affiliation{Department of Physics, The Hong Kong University of Science and Technology, Kowloon, Hong Kong, China}

\author{Wanyue Xiao}
\affiliation{Department of Physics, City University of Hong Kong, Tat Chee Avenue, Kowloon, Hong Kong, China}

\author{Jensen Li} 
\email{j.li13@exeter.ac.uk} 
\affiliation{Department of Physics, The Hong Kong University of Science and Technology, Kowloon, Hong Kong, China}
\affiliation{Centre for Metamaterial Research and Innovation, Department of Physics and Astronomy, University of Exeter, Exeter, United Kingdom}

\author{Shubo Wang} 
\email{shubwang@cityu.edu.hk} 
\affiliation{Department of Physics, City University of Hong Kong, Tat Chee Avenue, Kowloon, Hong Kong, China}

\begin{abstract}
\textbf{Real-space singularities govern a broad spectrum of wave phenomena, yet they remain largely unexplored in elastic wave systems. Here, we report hybrid topological singularities that emerge on the surfaces of finite-sized solids due to the full vectorial character of elastic waves. These textures fuse spin-field singularities with displacement-field singularities and exhibit unique non-pairwise topological charge dynamics. Moreover, a subset of these singularities imprint dislocation lines onto the radiated acoustic field, generating robust acoustic vortices in free space from an otherwise achiral source and structure. Our results establish elastic waves as a powerful platform for engineering real-space singularities and open avenues for singular phononics and the exploration of rich topological defects in elastic media.}
\end{abstract}

\maketitle


Wave singularities—spatial loci where wave properties are undefined—lie at the heart of wave field topology. In recent years, real-space wave singularities have attracted growing interest in both quantum and classical systems \cite{1uchida2010generation,2shen2019optical,3ni2021multidimensional,4li2024chip}, bridging topological physics and structured wave fields. Scalar waves (e.g., acoustic waves) can host phase singularities with undefined phase \cite{5nye1974dislocations}, while vector waves (e.g., electromagnetic waves) can support polarization singularities (e.g., C points and V points) with indeterminate polarization \cite{6nye1983lines,7hajnal1987singularities} and spin defects with ill-defined spin \cite{8wang2022topological,  9fang2025topological}. These real-space singularities are characterized by topological indices \cite{6nye1983lines,8berry2004index,9bliokh2019geometric} and give rise to exotic wave-field structures, including vortex knots \cite{10dennis2010isolated,11PhysRevLett.111.150404}, Möbius strips \cite{12freund2005cones,13freund2010optical,14bauer2015observation}, and skyrmions \cite{15tsesses2018optical,16du2019deep,17ge2021observation,18mata2025skyrmionic}. Their distinctive properties have enabled applications ranging from high-capacity communications \cite{19wang2012terabit,20shi2017high}, and optical/acoustic tweezers  \cite{21garces2003observation,22volke2008transfer} to super-resolution imaging  \cite{23hell1994breaking,24balzarotti2017nanometer}. Interactions between singularity-bearing waves and matter further generate intriguing spin-orbit interactions \cite{25marrucci2006optical,26wang2021spin,27xiao2026acoustic} and enable novel wave manipulation through the geometric topology of structures \cite{28peng2022topological,29fu2024near}.

Classical wave systems offer convenient platforms for exploring singular wave physics, owing to their experimental accessibility and the possibility of direct field observation. Early studies focused primarily on the singularities in electromagnetic waves  \cite{30dennis2009singular,31chen2019singularities,32liu2021topological,33zhou2024selective,34silveirinha2026thermodynamic}. More recently, investigations have been extended to acoustic  \cite{37muelas2022observation,36tong2025sculpturing} and water waves  \cite{39bliokh2021polarization, 40smirnova2024water, 38Wang2025,39Xiao2025}, uncovering rich phenomena rooted in the topological nature of real-space singularities.

Elastic waves span a broad range of scales, from macroscopic continuum mechanics to microscopic lattice vibrations \cite{40auld1973acoustic}. Their full vectorial character and coupling with other wave types make them a fertile ground for complex wave phenomena. Despite extensive research on the fundamental properties of elastic waves  \cite{41Long2018, 46bliokh2022elastic, 42Chaplain2022,44PhysRevLett.131.136102,45Zhao2023,46Yang2024}, the real-space topological singularities in elastic wave systems remain largely unexplored. 

In this work, we report hybrid topological singularities, termed H points, of elastic waves supported by finite-sized solids. The full vectorial character of elastic waves endows these H points with two distinct facets: they simultaneously manifest as spin singularities and displacement-field singularities. \textcolor{black}{Consequently, each H point carries a pair of topological charges associated with the spin field and displacement field, respectively.} We show that the topological phase transitions of H points go beyond the traditional pairwise generation and annihilation, exhibiting unique evolutions due to their hybrid nature. Moreover, we show that the H points can imprint dislocation lines onto the surrounding air, thereby generating acoustic vortices in free space from an achiral structure driven by a simple point source. This effect originates from the coupling between the vectorial elastic field and the scalar acoustic field, which projects the elastic field topology onto the radiative channel. Our results open a pathway toward exploring richer topological singularities arising from the full complexity of elastic waves.

We consider a spherical shell made of an isotropic solid material with radius $r$ and thickness $ t$, as illustrated in Fig. 1(a). The shell supports elastic waves with a complex vectorial displacement field $\mathbf{u}(\mathbf{r})$. This field is elliptically polarized in general and carries a spin density $\mathbf{S}=\text{Im} [\mathbf{u}^* \times \mathbf{u}]$ \cite{50yuan2021observation}. Our interest lies in the topological singularities of the displacement field $\mathbf{u}$ and the spin field $\mathbf{S}$, and we will focus on the singularities that couple with acoustic radiation channels in free space. These include the phase singularities of the flexural field $u_{\perp}$(i.e., displacement field normal to the surface) and the polarization singularities of the tangent spin field $\mathbf{S}_{\parallel}$ (i.e., spin field parallel to the surface). The tangent displacement field and the normal spin field cannot couple with free-space radiation channels, and their singularities will not be considered here. 

We carry out numerical simulations of the elastic waves supported by the spherical shell by using a finite-element method package, COMSOL. Without loss of generality, we set the shell material to be photosensitive resin with density $\rho = 1190  \mathrm{~kg} / \mathrm{m}^3$, Young’s modulus $E$ = 3.4 + 0.136i GPa, and Poisson ratio $\nu$ = 0.35. The shell is excited by two point sources $\mathrm{E}_1$ and $\mathrm{E}_2$ located on the surface, as marked by the red dots in Fig. 1(a). The two sources vibrate along the normal direction of the surface at a frequency of $f$ = 2100 Hz, with a phase difference of $\pi/2$. The detailed design principle for the system is provided in the Supplemental Material \cite{48supp_material}. 

\begin{figure}\label{Fig1}
\centering
\includegraphics[width=0.7\linewidth]{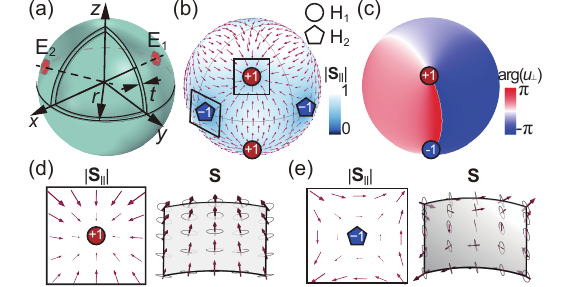} 
\caption{(a) Schematic of the solid shell. The shell has radius $r$ = 8 cm and thickness $t$ = 0.3 cm, and is excited by two sources $\mathrm{E}_1$ at ($\theta_1=90^{\circ}, \phi_1=180^{\circ}$) and $\mathrm{E}_2$ at ($\theta_1=90^{\circ}, \phi_1=290^{\circ}$). (b) Tangent spin field $\mathbf{S}_{\parallel}$ on the shell surface at the frequency $f$= 2100 Hz, excited by two point sources $\mathrm{E}_1$ and $\mathrm{E}_2$ with $\pi$/2 phase difference. (c) Phase of the corresponding flexural displacement field $u_{\perp}$. Distributions of spin vectors and polarization ellipses near the (d) $\mathrm{H}_1$ and (e) $\mathrm{H}_2$ points.}
\end{figure}

Figures 1(b) shows the simulated distributions of the tangent spin $\mathbf{S}_{\parallel}$ on the spherical surface. We notice that multiple singularities emerge in the tangent spin field. These singularities can be classified into two types: $\mathrm{H}_1$ points (marked as circles) and $\mathrm{H}_2$ points (marked as pentagons). At the $\mathrm{H}_1$ points, $\mathbf{S}_{\parallel}$ has zero magnitude with ill-defined direction [Fig. 1(d)], and $u_{\perp}$ has zero amplitude with ill-defined phase [Fig. 1(c)]. Thus, the $\mathrm{H}_1$ points are hybrid singularities with two facets: V points of $\mathbf{S}_{\parallel}$ field and phase singularities of $u_{\perp}$ field. At the $\mathrm{H}_2$ points, $\mathbf{S}$ has zero magnitude with ill-defined direction, and $\mathbf{u}$ is linearly polarized with an ill-defined polarization ellipse [Fig. 1(e)]. Thus, the $\mathrm{H}_2$ points are also hybrid singularities with two facets: $\mathrm{V}$ points of $\mathbf{S}$ field and L points of $\mathbf{u}$ field. The dual facets of $\mathrm{H}_1$ and $\mathrm{H}_2$ are not accidental but a unique property arising from the full vectorial character of elastic waves confined in a two-dimensional surface. Naturally, each hybrid singularity is characterized by a pair of topological charges: For an $\mathrm{H}_1$ point, the spin singularity carries the charge $q_{\mathbf{s}}=\frac{1}{2 \pi} \oint \nabla \varphi \cdot \mathrm{~d} \mathbf{r}$ ($\varphi$ is the local orientation angle of $\mathbf{S}_{\parallel}$), while the displacement-field singularity carries the charge $q_{u} =\frac{1}{2 \pi} \oint \nabla \operatorname{arg}\left(u_{\perp}\right) \cdot \mathrm{d} \mathbf{r}$; For an $\mathrm{H}_2$ point, the spin singularity carries the charge $q_{\mathbf{s}} = \frac{1}{2 \pi} \oint \nabla \varphi \cdot\mathrm{~d}\mathbf{r}$, while the displacement-field singularity carries the charge $q_\mathbf{u}= \frac{1}{2 \pi} \oint \nabla \varphi\cdot\mathrm{~d} \mathbf{r}$. Here, \textcolor{black}{the integrals are carried out over a small loop enclosing the singularities}; $q_\mathbf{u}=q_\mathbf{s}$ for $\mathrm{H}{_2}$ because the spin is always perpendicular to the minor axis of the polarization ellipse \cite{7hajnal1987singularities, 8berry2004index}.  These topological charges of $\mathrm{H}_1$ and $\mathrm{H}_2$ points are labeled in Figs. 1(b) and 1(c).

The hybrid singularities $\mathrm{H}_1$ and $\mathrm{H}_2$ possess intriguing hybrid global topological properties. First, the sum of $q_{\mathbf{s}}$ of all the spin singularities on the spherical surface is equal to the Euler characteristic of sphere, which is a direct result of the Poincaré–Hopf theorem applied to vector fields (i.e., $\mathbf{S}_{\parallel}$) on a smooth manifold (i.e., spherical surface) \cite{53needham2021visual}. For the case in Fig. 1(b), in total there are six $\mathrm{H}_1$ singularities with $q_{\mathbf{s}} = +1$ and four $\mathrm{H}_2$ singularities with $q_{\mathbf{s}} = -1$ on the surface. The sum of their topological charges is indeed $\sum q_{\mathbf{s}} = 2$, agreeing with the Euler characteristic of sphere. Second, the sum of $q_u$ of all the flexural field singularities on the spherical surface is always zero, which is guaranteed by the continuity of the flexural field \cite{51Fu2024}. For the case in Fig. 1(c), in total there are three singularities with $q_u = +1$ and three singularities with $q_u = -1$ on the surface. The total topological charge is indeed $\sum q_u = 0$. Notably, these singularity configurations (types, number, charge, and location) are governed by the modal superposition that depends on both the solid geometry and the excitation \cite{48supp_material}, yet their global topological properties (i.e., total charge) are robust against continuous deformations of the solid and independent of the excitation properties.

We employ a scanning laser Doppler vibrometer (LDV) to measure the flexural displacement field $u_{\perp}$ on the spherical surface, as illustrated in Fig. 2(a). The spherical shell is fabricated via 3D printing using cured photosensitive resin. Two piezoelectric transducers attached to the shell, driven by a waveform generator, serve as point sources $\mathrm{E}_1$ and $\mathrm{E}_1$, exciting normal vibrations on the surface. The complex flexural field $u_{\perp}$ is mapped over the regions enclosed by the solid lines in Fig. 1(b). The measured amplitude and phase distributions in these two regions are shown in Figs. 2(d) and 2(e). Figure 2(e) shows a vanishing amplitude accompanied by a phase singularity, in agreement with the numerical predictions in Fig. 2(b). A quantitative analysis of the phase winding of the flexural field around $\mathrm{H}_1$ points is given in \cite{48supp_material}. In contrast, for the $\mathrm{H}_2$ point, the measured flexural field [Fig. 2(e)] exhibits a regular amplitude and a smooth phase profile, with no indication of phase singularity, in agreement with the simulations [Fig. 2(c)].  

\begin{figure}\label{Fig2}
\centering
\includegraphics[width=0.7\linewidth]{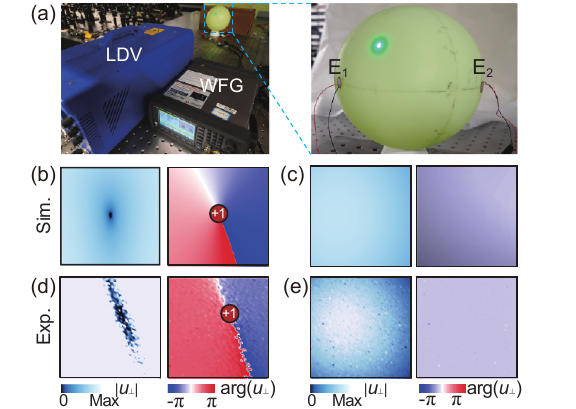} 
\caption{(a) Experimental setup for mapping the displacement field of flexural waves on the shell, including a scanning laser Doppler vibrometer (LDV) and a wave function generator (WFG). The inset shows a zoom-in of the sample with two piezoelectric transducers ($\mathrm{E}_1$ and $\mathrm{E}_2$) serving as point sources to excite elastic waves. The green dot is the laser spot. Numerically simulated amplitude and phase of flexural field $u_{\perp}$ near the (b) $\mathrm{H}_1$ point and (c) $\mathrm{H}_2$ point marked in Fig. 1(b). Experimentally measured amplitude and phase of flexural field $u_{\perp}$ near the (d) $\mathrm{H}_1$ point and (e) $\mathrm{H}_2$ point marked in Fig. 1(b). All system parameters are the same as those in Fig. 1.}
\end{figure}

The hybrid topological singularities undergo evolutions under parameter variations and can give rise to nontrivial topological transitions, as illustrated in Fig. 3 for three example cases. The topological phase transition here refers to the generation and annihilation of hybrid singularities, which should be distinguished from the transition between distinct topological phases of matter \cite{55ozawa2019topological}. The evolutions of the hybrid singularities are governed by two fundamental constraints: (i) the total topological charge of the hybrid singularities is conserved, and (ii) direct annihilation can only happen to the singularities of the same type, i.e., $\mathrm{H}_1$ ($\mathrm{H}_2$) points can annihilate with $\mathrm{H}_1$ ($\mathrm{H}_2$) points only. The second constraint is attributed to the different topological natures of $\mathrm{H}_1$ and $\mathrm{H}_2$ points, i.e., they correspond to the hybridization of different types of scalar/vector singularities. Figure 3(a) shows the topological phase transition for a pair of $\mathrm{H}_1$ points under varying excitation frequency, where the spin-field charge $q_{\mathbf{s}} = +1$ annihilates with $q_{\mathbf{s}} = -1$ and the flexural-field charge $q_u = +1$ annihilates with $q_u = -1$. Figure 3(b) shows the topological phase transition for a pair of $\mathrm{H}_2$ points under varying excitation frequency. At the $\mathrm{H}_2$ points, the charge of the spin-field V point is always equal to the charge of the displacement-field L point (i.e.,$q_{\mathbf{s}} = q_{\mathbf{u}}$), and flexural field has no phase singularity. Consequently, we observe the annihilation of $q_{\mathbf{s}} = +1$ and $q_{\mathbf{s}} = -1$ in Fig. 3(b).

\begin{figure}\label{Fig3}
\centering
\includegraphics[width=0.7\linewidth]{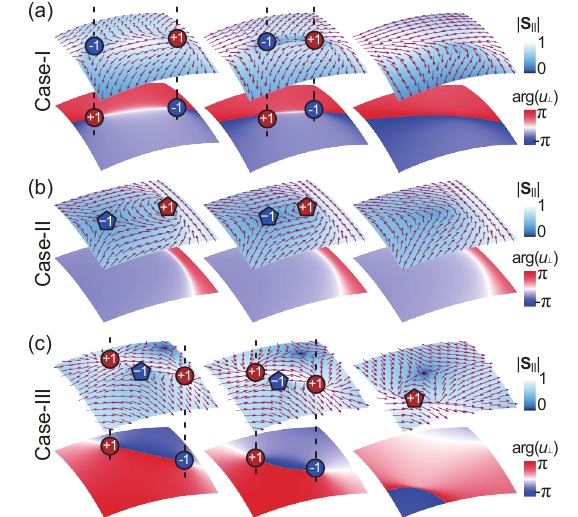} 
\caption{Topological phase transitions of the hybrid singularities. (a) Annihilation of two $\mathrm{H}_1$ points with opposite charges. (b) Annihilation of two $\mathrm{H}_2$ points with opposite charges. (c) Topological phase transition of two $\mathrm{H}_1$ points and one $\mathrm{H}_2$ point. The upper panels show the distributions of tangent spin field $\mathbf{S}_{\parallel}$ with the background show the spin magnitude $|\mathbf{S}_{\parallel}|$, while the lower panels show the phase distributions of the flexural field $u_{\perp}$. All the geometric parameters are the same as those in Fig. 1. The varied parameter is the frequency $f$: (a) $f$ = 3246 Hz, 3248 Hz, and 3250 Hz; (b). $f$ = 3520 Hz, 3540 Hz, and 3550 Hz; (c) $f$ = 3700 Hz, 3630 Hz, and 3533 Hz.}
\end{figure}

A notable topological phase transition is shown in Fig. 3(c), which involves two $\mathrm{H}_1$ points and one $\mathrm{H}_2$ point. In this case, the two $\mathrm{H}_1$ points do not form a topological pair because they have the same spin-field charges of $q_{\mathbf{s}} = +1$. Remarkably, they still can annihilate with each other through a “topological charge transfer” mechanism, where they transfer their spin-field charges to the $\mathrm{H}_2$ point with $q_{\mathbf{s}} = -1$, converting it to an $\mathrm{H}_2$ point with $q_{\mathbf{s}} = +1$. During this process, the total topological charge of $\mathrm{H}_1$ and $\mathrm{H}_2$ remains conserved. From the perspective of the displacement field [lower panel of Fig. 3(c)], the two $\mathrm{H}_1$ points carry opposite topological charges  $q_u = +1$ and $q_u = -1$ and are annihilated with each other. This unique non-pairwise topological charge dynamics is attributed to the distinct topological constituents of $\mathrm{H}_1$ and $\mathrm{H}_2$ points, highlighting the essential role of the vectorial degrees of freedom in elastic waves.

\begin{figure}\label{Fig4}
\centering
\includegraphics[width=0.7\linewidth]{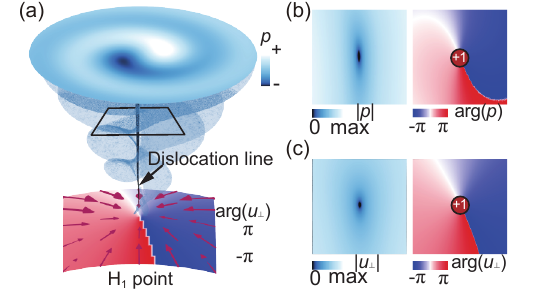} 
\caption{(a) Acoustic vortex generation by the $\mathrm{H}_1$ point marked in Fig. 1(c). (b) The amplitude and phase distribution of the pressure field p in the region enclosed by the black box in (a). (c) The amplitude and phase distribution of the flexural field $u_{\perp}$ near the $\mathrm{H}_1$ point in (a).}
\end{figure}

Since the $\mathrm{H}_1$ points carry phase singularities of the displacement field $u_{\perp}$, their topological characteristics can be transferred to the surrounding acoustic fields in free space via elastic-acoustic coupling. Consequently, each $\mathrm{H}_1$ point on the shell generates a dislocation line in the radiated pressure field, as shown in Fig. 4(a). Around these lines, the acoustic field exhibits the characteristic features of a vortex beam. A representative cross-section (enclosed by the black square in Fig. 4(a)) of the pressure field is shown in Fig. 4(b). We clearly see that the pressure amplitude approaches zero at the center (on the dislocation line), accompanied by a well-defined phase vortex. Importantly, the topological charge of the acoustic vortex is the same as that of the flexural field $u_{\perp}$ on the shell surface, as shown in Fig. 4(c). The mode purity of the acoustic vortex is about 71\% in the current case, and it can be generated over a wide frequency range of 1300–2500 Hz, corresponding to a fractional bandwidth of approximately 63\% \cite{48supp_material}. This provides a simple yet robust mechanism for generating acoustic vortices in free space using a minimal setup. Conventional methods of acoustic vortex generation rely on prescribed phase arrangements, such as spiral phase plates \cite{56wang2016particle}, phased arrays \cite{57ye2016making}, and resonators \cite{58jiang2016convert}, which require precise phase control and involve complex structural designs, thereby being sensitive to perturbations. In contrast, the present mechanism originates from the topological singularities of elastic waves and is robust to small perturbations. We note that the $\mathrm{H}_1$ points can also generate acoustic vortices inside the shell through elastic-acoustic coupling \cite{48supp_material}. Furthermore, higher-order acoustic vortices can be realized based on higher-order singularities of the flexural field, which may arise under additional symmetry (e.g., rotational symmetry) \cite{48supp_material}, offering further flexibility for structured wave generation.

\begin{figure}\label{Fig5}
\centering
\includegraphics[width=0.7\linewidth]{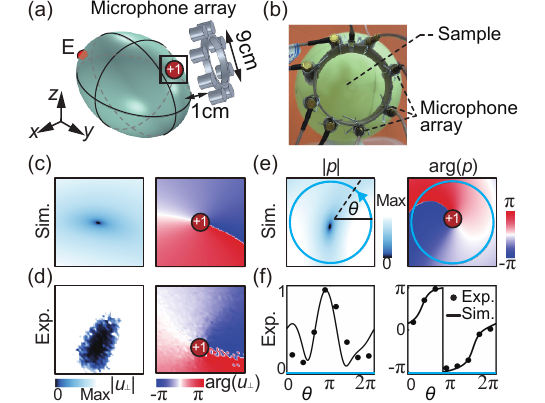} 
\caption{(a) Schematic for acoustic vortex generation by an ellipsoidal shell driven by a point source. (b) A photo of the microphone array used to measure the pressure field. Simulated (c) and experimentally measured (d) amplitude and phase distribution of the flexural field near the $\mathrm{H}_1$ point marked by a solid box in (a). (e) Simulated pressure field in the measurement plane. (f) Experimentally measured pressure field along the blue circle in (e). The ellipsoid has radii of $r_y =1.05r_x=1.05r_z$ with $r_x$ = 7 cm, and the shell thickness is 0.3 cm. The piezoelectric transducer $\mathrm{E}$ has a radius of 0.5 cm and \textcolor{black}{located at ($\theta=80^{\circ}, \phi=285^{\circ}$)}. The operating frequency is 2620 Hz. }
\end{figure}

The above framework can be employed to realize a counterintuitive phenomenon—the generation of acoustic vortices from an achiral structure driven by a single, non-chiral source. As shown in Fig. 5(a), we consider an ellipsoidal shell excited by a point source $\mathrm{E}$. The resulting elastic field, governed by the interference of the excited eigenmodes \cite{48supp_material}, exhibits multiple hybrid singularities. Focusing on the $\mathrm{H}_1$ point marked in Fig. 5(a), both simulations and measurements confirm the presence of a vanishing flexural amplitude and a phase singularity of charge $q_u = +1$, as shown in Figs. 5(c) and 5(d). As established, this $\mathrm{H}_1$ point is able to generate an acoustic vortex in free space, as shown in Fig. 5(e), where the simulated pressure field exhibits a clear singularity and phase winding pattern. To experimentally verify this vortex generation, we employ a circular microphone array (BK 4958) positioned 1 cm away from the $\mathrm{H}_1$ point, with a diameter of 9 cm, as illustrated in Fig. 5(a, b). Figure 5(f) shows the simulation (solid lines) and experimental (dots) results for the pressure amplitude and phase along the circular array. We clearly notice the 2$\pi$ phase winding, with an excellent agreement between the simulation and experimental results, providing direct evidence of the radiated acoustic vortex. These results demonstrate that $\mathrm{H}_1$ points can give rise to chiral elastic–acoustic responses without relying on chiral geometries or sources. This highlights a fundamentally different route to structured wave generation, where chirality emerges from structural topology rather than symmetry. 

We have demonstrated the emergence and evolution of real-space hybrid topological singularities of elastic waves in finite-sized solids. These singularities arise from the full vectorial character of elastic fields constrained to a two-dimensional surface and exhibit nontrivial topological dynamics, including non-pairwise transitions that go beyond the conventional defect dynamics in two-dimensional systems. More importantly, we show that a subset of these singularities, $\mathrm{H}_1$ points, imprint phase singularities onto the radiated acoustic field, enabling the generation of a vortex in free space. This mechanism leads to chiral acoustic radiation from achiral structures driven by simple achiral sources, thereby establishing a topology-driven route to structured wave generation. Our findings bridge the gap between the intrinsic wave topology in elastic media and observable wave phenomena, opening new avenues for singular phononics and advanced topological wave manipulation.

The work described in this paper was supported by grants from the National Natural Science Foundation of China (No. 12322416), the Research Grants Council of the Hong Kong Special Administrative Region, China (Project No. AoE/P-502/20), and the Talent Scientific Fund of Lanzhou University. We thank C. T. Chan and Chenwen Yang for helpful discussions.

\textit{Data availability}---The data that support the findings of this article are not publicly available. The data are available from the authors upon reasonable request.


\bibliography{apssamp}
\bibliographystyle{apsrev4-2}

%


\newpage


\setcounter{figure}{0}
\setcounter{table}{0}
\setcounter{equation}{0}
\setcounter{page}{1} 


\pagebreak
\widetext
\begin{center}
\textbf{\large Supplementary Materials for \\
Real-space Hybrid Topological Singularities in Structured Elastic Waves}\\
Tong Fu$^{ 1}$, Pengfei Zhao$^{ 2}$, Liyou Luo$^{2}$, Zhiling Zhou$^{1}$,
Dong Liu$^{2}$, Wanyue Xiao$^{1}$, \\ Jensen Li$^{2,3}$, and Shubo Wang$^{1}$\\ 

$^{1}$\textit{Department of Physics, City University of Hong Kong, Tat Chee Avenue, Kowloon, Hong Kong, China}\\

$^{2}$\textit{Department of Physics, The Hong Kong University of Science and Technology, Kowloon, Hong Kong, China}\\
$^{3}$\textit{Centre for Metamaterial Research and Innovation, Department of Physics and Astronomy, University of Exeter, Exeter, United Kingdom}\\

\end{center}

\tableofcontents
\newpage

\section{NOTE 1. Elastic experimental details}
The experimental setup consists of a scanning laser Doppler vibrometer (LDV, OptoMET) and a waveform generator (WFG, Keysight 33500B). The LDV laser was directed perpendicular to the sample surface to measure the out-of-plane velocity distribution. The two samples (spherical and ellipsoidal shells) were fabricated via 3D printing (iSLA800) using a cured photosensitive resin with density $\rho = 1190  \mathrm{~kg} / \mathrm{m}^3$, Young’s modulus $E$ = 3.4 + 0.136i GPa, and Poisson ratio $\nu$ = 0.35. 

The spherical shell has a radius $r$ = 8 cm and a uniform thickness $t$ = 0.3 cm (Figs. 1-4 in the main text). Excitation was implemented using two piezoelectric transducers  E$_1$ and E$_2$ (radius 0.55 cm), positioned in the $xy$-plane at ($\theta_1=90^{\circ}, \phi_1=180^{\circ}$) and ($\theta_2=90^{\circ}, \phi_2=290^{\circ}$), respectively. The transducers were driven by the WFG with a constant amplitude of 9 V; Channel 1 drove E$_1$ with a 0 phase, while Channel 2 drove E$_2$ with a $\pi$/2 phase delay. The ellipsoidal shell has a thickness $t$ = 0.3 cm and radii $r_y =1.05r_x$ with $r_x=r_z$ = 7 cm (Fig. 5 in the main text). For this geometry, only a single piezoelectric transducer E was used as the point source. 

The schematic of the experimental setup, including the LDV, sample, waveform generator, and data-acquisition system, is shown in Fig. S1. Flexural wave propagation is measured by non-contact optical detection of surface motion via the Doppler effect. A coherent laser beam is focused onto the surface of the curved elastic continuum, where local vibrations—induced by piezoelectric excitation—produce a frequency shift (i.e., the Doppler shift) in the backscattered light. This shift is directly proportional to the instantaneous out-of-plane surface velocity. Using a heterodyne interferometer, the scattered light is mixed with an internal reference beam that is frequency-shifted by a known amount, enabling demodulation of the Doppler signal and accurate reconstruction of the vertical velocity field distribution on the spherical and ellipsoidal shells. 

\begin{figure}[htb]
\renewcommand{\figurename}{FIG.}
\renewcommand{\thefigure}{S\arabic{figure}}
\centering
\includegraphics[width=0.7\linewidth]{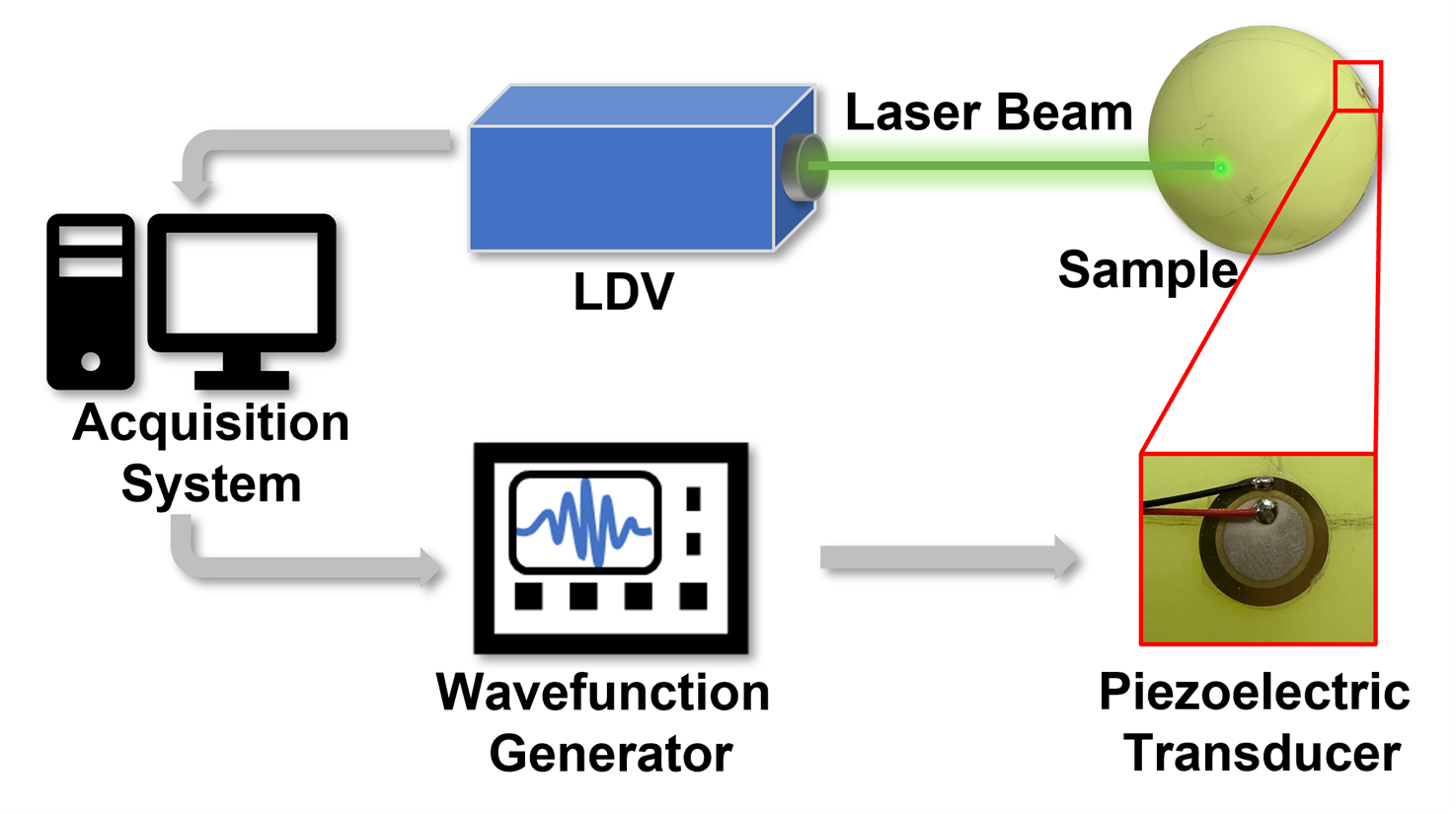} 
\caption{Schematic of the experimental setup for generating and measuring out-of-plane velocity (flexural fields) distribution.}
\end{figure}

\newpage


\section{NOTE 2. Design principles of hybrid singularities}

The emergence of the elastic-wave hybrid singularities is guaranteed by the Poincaré–Hopf theorem. As long as the solid has a smooth surface, such singularities must emerge on its surface under arbitrary excitation. The specific configuration of these singularities (types, charges, locations, etc.) depends on the detailed settings, i.e., shell geometry and source positions/phases, which determine the excited elastic eigenmodes and their interference. The singularity pattern results from the complex modal superposition. More specifically, the elastic field on the shell can be regarded as a superposition of eigenmodes,
$$
\mathbf{u}(\mathbf{r}, \omega)=\sum_n a_n(\omega) \Psi_n(\mathbf{r})
$$
where the modal coefficient $a_n$ is determined by the coupling between the source (E) and the $n$-th eigenmode. This coupling depends on the source position relative to the modal profile, the source phase, and the detuning from the modal frequency. Therefore, changing the shell geometry modifies the modal basis $\Psi_n$, and changing the source configuration modifies the complex modal weights $a_n$. Together, they determine the interference displacement field $\mathbf{u}$ and thus the singularity pattern. Nevertheless, two general principles and constraints guide the design of singularity configurations.

\vspace{1em}

\noindent\textbf{1. Global topological constraint}

Although the exact number and positions of the singularities depend on the detailed setting (shell geometry, source positions/phases, etc.), the total topological charge $q_{{\mathbf{s}}}$ of all the hybrid singularities on the shell surface, is constrained by the Poincaré–Hopf theorem. For a closed shell with smooth surfaces, the sum of the topological charges of all the singularities must satisfy $\sum q_{\mathbf{s}}$ = $\chi$, where $\chi$ is the Euler characteristic of the surface. In the revised manuscript, both the spherical and the ellipsoidal shells have an Euler characteristic of $\chi = 2$. As a result, the sum of the topological charges $q_{\mathbf{s}}$ is always 2. While the singularities may move, split, merge, or appear in different locations depending on the specific modal superposition, their total charge remains constant. This provides a global constraint for engineering the singularities on the shells.

\begin{figure}[htb]
\renewcommand{\figurename}{FIG.}
\renewcommand{\thefigure}{S\arabic{figure}}
\centering
\includegraphics[width=0.7\linewidth]{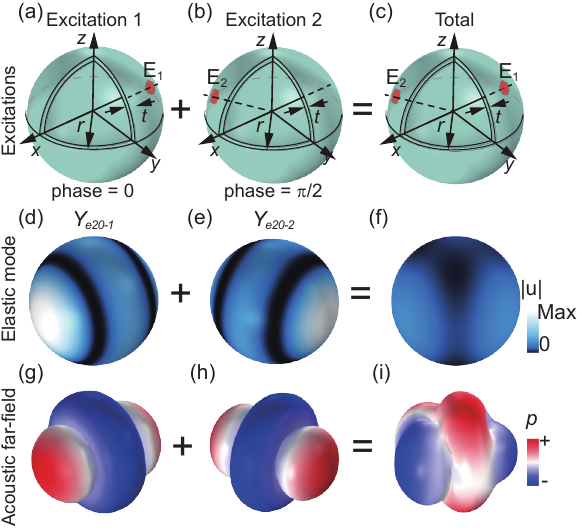}
\caption{\textbf{Modal superposition in the spherical shell.}
(a) Schematic of excitation with only $\mathrm{E}_1$, located at 
$(\theta_1 = 90^{\circ}, \phi_1 = 180^{\circ})$, with the phase of excitation set to 0.
(b) Schematic of excitation with only $\mathrm{E}_2$, located at 
$(\theta_2 = 90^{\circ}, \phi_2 = 290^{\circ})$, with the phase of excitation set to $\pi/2$.
(c) Schematic of combination excitation with $\mathrm{E}_1$ at 
$(\theta_1 = 90^{\circ}, \phi_1 = 180^{\circ})$ and $\mathrm{E}_2$ at 
$(\theta_2 = 90^{\circ}, \phi_2 = 290^{\circ})$, with an excitation phase difference of $\pi/2$.
(d) Elastic eigenmode $Y_{e20\text{-}1}$ excited in (a).
(e) Elastic eigenmode $Y_{e20\text{-}2}$ excited in (b).
(f) Total elastic field distribution in (c). (g) The pressure far-field corresponding to the eigenmode in (d). (h) The pressure far-field corresponding to the eigenmode in (e)
(i) Total pressure far field corresponding to the combined excitation setup in (c).}
\label{fig:S2}
\end{figure}

\vspace{1em}

\noindent\textbf{2. Modal superposition principle}

While the emergence of in-plane spin singularities are protected by geometric topology of the solids, the emergence of isolated $\mathrm{H}_1$ points requires the superposition of at least two elastic eigenmodes with a nonzero relative phase. A single eigenmode, or a purely in-phase (with relative phase 0) or out-of-phase (with relative phase $\pi$) superposition of eigenmodes, generally does not produce isolated hybrid singularities or flexural-wave singularities. Specifically, the cases in Figs. 1–4 of the main text can be understood as induced by the out-of-phase modal superposition, as illustrated in Fig. S2. The two sources have a phase difference of $\pi$/2 and are located at different positions on the shell, predominantly exciting two orthogonal eigenmodes, $Y_{e20-1}$ and $Y_{e20-2}$, respectively. The mode-labeling convention and the meaning of the subscripts are detailed in Note 5. The interference of the two eigenmodes generates the complex elastic field in Fig. S2(f). The corresponding radiation pressure far-fields are shown in Figs. S2(g-i). The zeros of the pressure field in Fig. S2(i) correspond to the acoustic vortices. 

The above principle also applies to the ellipsoidal shell in Fig. 5 of the main text, which involves only one point source. As illustrated in Fig. S3, the ellipsoidal geometry breaks the degeneracy between $Y_{e20}$ and $Y_{e22}$ modes. Please refer to Note 5 for multipole expansions. Under the excitation of the point source E, the two modes are excited simultaneously. In the frequency range of 2560–2660 Hz, these two modes dominate the response and have different resonance frequencies. At the operating frequency of 2620 Hz, the $Y_{e22}$ mode is near resonance, whereas the $Y_{e20}$ mode is detuned. This detuning leads to a phase difference of approximately $\pi$/2 between the two modes. Their interference gives rise to the hybrid singularities in Fig. S3(a).

\begin{figure}[htb]
\renewcommand{\figurename}{FIG.}
\renewcommand{\thefigure}{S\arabic{figure}}
\centering
\includegraphics[width=0.7\linewidth]{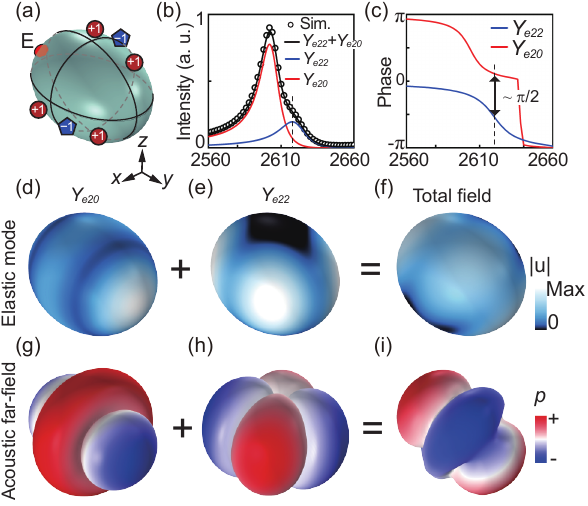}
\caption{\textbf{Modal superposition in the ellipsoidal shell.}
(a) Schematic of the excitation by a single source $\mathrm{E}$ located at 
$(\theta = 80^{\circ}, \phi = 285^{\circ})$.
(b) Acoustic field radiation spectra and corresponding multipole decomposition.
(c) Phase spectra of the relevant multipole modes.
(d) The elastic eigenmode $Y_{e20}$.
(e) The elastic eigenmode $Y_{e22}$.
(f) Total elastic field distribution excited by the source configuration in (a).
The far-field acoustic field emission pattern of (g) $Y_{e20}$, 
(h) $Y_{e22}$, and (i) total far-field acoustic radiation pattern resulting from their superposition.}
\label{fig:S3}
\end{figure}

More broadly, this provides a systematic design strategy for engineering the singularities on curved elastic shells. First, the shell geometry can be designed to tailor the modal spectrum, for example, by preserving, splitting, or shifting modal degeneracies. Second, source positions can be chosen to enhance coupling to desired modes while suppressing coupling to undesired modes. Third, the relative source phases and the operating frequency can be tuned to control the complex modal weights. Finally, the resulting spin field can be evaluated to locate its zeros and determine their topological charges, subject to the global constraint imposed by the Euler characteristic of the structure. This modal-superposition framework therefore provides a practical route to designing singularity configurations with desired numbers, locations, and charge distributions, although the final pattern generally requires solving the full modal interference problem rather than following a simple closed-form rule.

\newpage

\section{NOTE 3. Quantitative analysis of phase winding number near $\mathrm{H}_1$}

\begin{figure}[htb]
\renewcommand{\figurename}{FIG.}
\renewcommand{\thefigure}{S\arabic{figure}}
\centering
\includegraphics[width=0.6\linewidth]{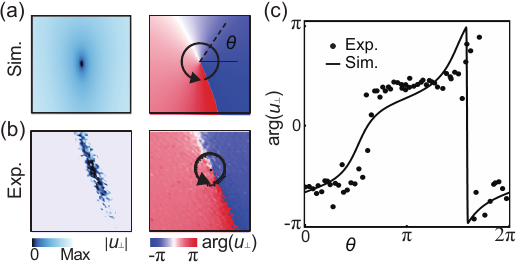}
\caption{\textbf{Quantitative analysis of the flexural wave phase winding of the $\mathrm{H}_1$ singularity on the spherical shell.}
The simulated (a) and experimentally measured (b) intensity and phase of the flexural field $u_{\perp}$.
(c) Phase of the flexural field along the black circle in the right panels of (a) and (b).
The singularity corresponds to the $\mathrm{H}_1$ point in Fig.~2 of the main text.}
\label{fig:S4}
\end{figure}

To better illustrate the phase singularities of flexural fields around $\mathrm{H}_1$, a quantitative phase analysis is presented in the following figures. Figure S4 shows the results for the singularity on the spherical shell, corresponding to the $\mathrm{H}_1$ point in Fig. 2 of the main text. Figure S4(c) shows the line plot of the flexural wave phase along a closed contour encircling the singularity in Fig. S4(a), which clearly exhibits a quantized $2\pi$ phase accumulation, thereby confirming a phase singularity with a topological charge of $q_u = +1$. In addition, quantitative analyses of the singularities on the ellipsoidal shell are also presented in Figs. S5 and S6. As shown in Fig. S5(d), the phase variation along a closed contour encircling the singularity in Fig. S5(b) also shows a quantized $2\pi$ phase accumulation, corresponding to a phase singularity with a charge of $q_u = +1$. By contrast, Fig. S6 shows a phase accumulation of –2$\pi$, indicating a phase singularity with a charge of $q_u = -1$.

\begin{figure}[htb]
\renewcommand{\figurename}{FIG.}
\renewcommand{\thefigure}{S\arabic{figure}}
\centering
\includegraphics[width=0.6\linewidth]{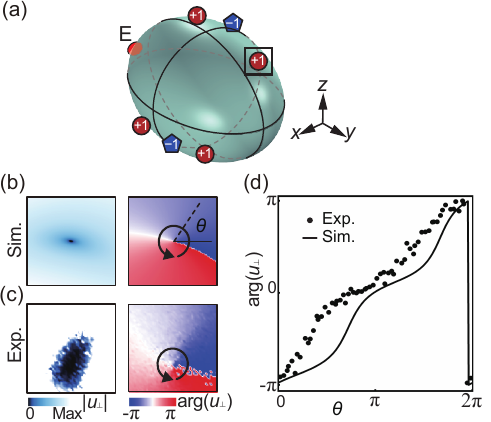}
\caption{\textbf{Quantitative analysis of the flexural wave phase winding of the $\mathrm{H}_1$ singularity on the ellipsoid shell.}
(a) Hybrid singularities on the ellipsoidal shell.
The simulated (b) and experimentally measured (c) intensity and phase of the flexural field $u_{\perp}$ near the singularity enclosed by the black square in (a).
(d) Phase of the flexural field along the black circle in the right panels of (a) and (b).
The singularity corresponds to the $\mathrm{H}_1$ point in Fig.~5 of the main text.}
\label{fig:S5}
\end{figure}

\begin{figure}[htb]
\renewcommand{\figurename}{FIG.}
\renewcommand{\thefigure}{S\arabic{figure}}
\centering
\includegraphics[width=0.6\linewidth]{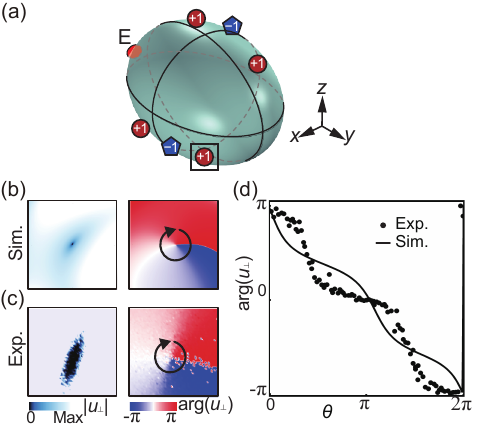}
\caption{\textbf{Quantitative analysis of the flexural wave phase winding of the $\mathrm{H}_1$ singularity on the ellipsoid shell.}
(a) Hybrid singularities on the ellipsoidal shell.
The simulated (b) and experimentally measured (c) intensity and phase of the flexural field $u_{\perp}$ near the singularity enclosed by the black square in (a).
(d) Phase of the flexural field along the black circle in the right panels of (a) and (b).
The parameters are the same as those in Fig.~5 of the main text.}
\label{fig:S6}
\end{figure}

\newpage

\section{NOTE 4. High-order singularities and acoustic vortices}

In the main text, the $\mathrm{H}_1$ exhibits a flexural-wave phase singularity that generates an acoustic vortex of the lowest order. In principle, when two flexural-wave singularities with the same topological charge collide, they can merge into a higher-order singularity. Such a higher-order singularity can imprint a higher-order phase dislocation onto the radiated acoustic field, thereby generating a higher-order acoustic vortex in free space. Generally, the emergence of a higher-order singularity requires certain spatial symmetry, such as rotational symmetry. An example is shown in Fig. S7(a, b), where the spherical shell is excited by eight point sources $\mathrm{E}_{1-8}$ evenly distributed on the equator. The sources have the same amplitude except for $\mathrm{E}_5$, which has an amplitude $A$. By varying $A$, we can observe the formation of a higher singularity.

\begin{figure}[htb]
\renewcommand{\figurename}{FIG.}
\renewcommand{\thefigure}{S\arabic{figure}}
\centering
\includegraphics[width=\linewidth]{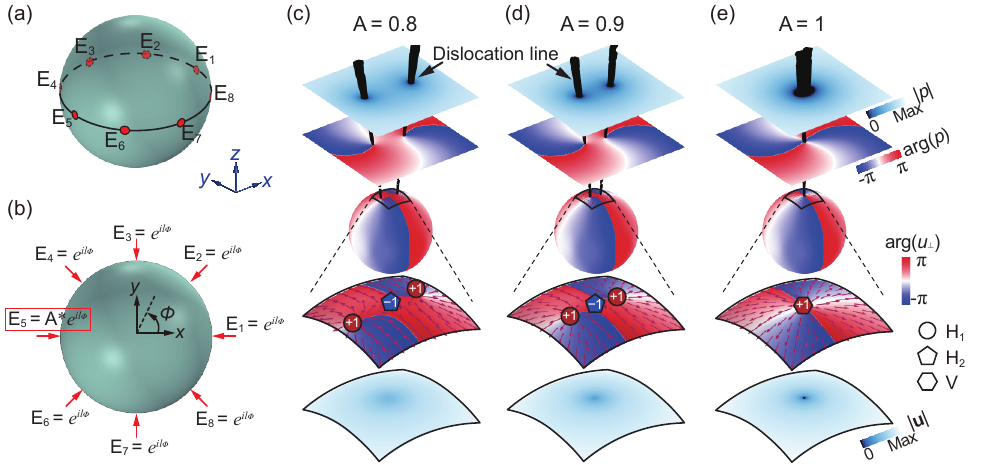}
\caption{\textbf{Generation of a high-order acoustic vortex.}
(a) Schematic of the system for generating an acoustic vortex.
The spherical shell is excited by eight sources evenly distributed on the equator.
(b) The vibration direction and phase of the sources.
The field and phase distributions for different amplitudes of $\mathrm{E}_5$:
(c) $A = 0.8$, (d) $A = 0.9$, and (e) $A = 1$ with rotational symmetry.
The geometric parameters and operating frequency are the same as those in Fig.~1 of the main text.}
\label{fig:S7}
\end{figure}

As shown in Figs. S7(a) and S7(b), the source phases are prescribed as $e^{il\phi}$ with $l \in \mathbb{Z}$ and $\phi$ being the azimuthal coordinate. Throughout the Supplementary Material, we use $l$ to denote the topological charge of acoustic vortices. Namely, the sources have an azimuthal phase increment of $2\pi l$. As shown in Fig.~S7(c--e), when the source amplitude $A$ approaches 1
(rotational symmetry), we observe the merging of two $\mathrm{H}_1$
singularities with charge $q_\mathbf{s} = +1$ and one $\mathrm{H}_2$ singularity
with charge $q_\mathbf{s} = -1$. The resulting singularity is a $\mathrm{V}$ point
of the displacement field $\mathbf{u}$ with $|\mathbf{u}| = 0$. The
corresponding flexural wave singularity is second order with $q_u = -2$,
as indicated by its phase distribution, showing a $-4\pi$ phase winding
around the singular point. This second-order flexural-wave singularity can
then radiate into free space, generating a second-order acoustic vortex
[upper panel of Fig.~S7(e)].

Therefore, higher-order acoustic vortices can, in principle, be generated
through the merging of same-charge flexural-wave singularities. In the
present case, this process is realized through a hybrid singularity
topological transition, in which the overlap of $\mathrm{H}_1$ and
$\mathrm{H}_2$ hybrid singularities is accompanied by the emergence of a
second-order flexural-wave phase singularity and the corresponding
second-order acoustic vortex.

\newpage

\section{NOTE 5. Multipole expansions}

The acoustic fields radiated from the shells can be decomposed into scalar spherical harmonics (multipoles). The complex amplitude of pressure field $p(\mathbf{r})$ can be written as\cite{tsimokha2022acoustic}
\begin{equation}
p(\mathbf{r})=i k \sum_{l=0}^{\infty} \sum_{m=-l}^l a_{\mathcal{P} l m} Y_{\mathcal{P} l m}(\theta, \phi) h_l^{(1)}(k r),
\end{equation}
where $k$ is the wave number, $a_{\mathcal{P} l m}$  are the complex coefficients of the multipoles, $Y_{\mathcal{P} l m}(\theta, \phi)$ is the complex spherical harmonics, $h_l^{(1)}(k r)$ is the spherical Hankel function of the first kind. The eigenmodes can be designated by scalar spherical harmonics $Y_{\mathcal{P} l m}(\theta, \phi)$, where the lower indices $\mathcal{P}$, $l$, and $m$ denote the parity symmetry, orbital angular quantum number, and azimuthal quantum number, respectively. And $\mathcal{P} = e, o$ designates the even and odd parity symmetry of the modes. In Fig. S3, the total radiation power ($I$) was obtained by integrating the energy flux over a surface surrounding the shells

\begin{equation}
I=\int_0^{2 \pi} \int_0^\pi \boldsymbol{\Pi} \cdot \widehat{\mathbf{n}} r^2 \sin \theta d \theta d \phi
\end{equation}
where $\boldsymbol{\Pi}=\frac{1}{2} \operatorname{Re}\left(p^* \mathbf{v}\right)$ denote the time-averaged energy flux density with $p$ being the pressure field and $\mathbf{v}$ being the velocity field, and $\hat{\mathbf{n}}$ is the outer normal direction of the chosen spherical surface for integration. For scatterers of arbitrary shapes, the coefficients $a_{\mathcal{P} l m}$ can be determined as,

\begin{equation}
a_{\mathcal{P} l m}=\frac{\int_0^{2 \pi} \int_0^\pi p(\theta, \phi) Y_{\mathcal{P} l m}^*(\theta, \phi) \sin \theta \mathrm{d} \theta \mathrm{~d} \phi}{i k h_l^{(1)}(k r) \int_0^{2 \pi} \int_0^\pi Y_{\mathcal{P} l m}^*(\theta, \varphi) Y_{\mathcal{P} l m}(\theta, \phi) \sin \theta \mathrm{d} \theta \mathrm{~d} \phi}
\end{equation}
And the phase shown in Fig. S3 is arg($a_{\mathcal{P} l m}$).

\newpage

\section{NOTE 6. Quantitative characterization of acoustic orbital angular momentum}

\noindent\textbf{1.~OAM mode purity}

Theoretically, the acoustic vortex modes of different orders are orthogonal.
Therefore, the generated acoustic vortex can be decomposed into a superposition of vortex modes as \cite{li2024tunable}

\begin{equation}
p(r,\phi,z)
=
\sum_{l=-\infty}^{\infty}
\frac{1}{\sqrt{2\pi}} b_l(r,z)\exp(il\phi),
\end{equation}
where $p$ denotes the pressure field of the generated acoustic vortex. Here,
$r$, $\phi$, and $z$ are the radius, azimuthal angle, and height in the cylindrical coordinate system, respectively.
The coordinate system is defined with the $z$-axis aligned with the orbital
angular momentum emission direction, rather than the laboratory coordinate
system. The coefficient of the $l$-th order vortex mode can be written as
\begin{equation}
b_l(r,z)
=
\frac{1}{\sqrt{2\pi}}
\int_{0}^{2\pi}
p(r,\phi,z)\exp(-il\phi)\,d\phi .
\end{equation}

The power carried by each acoustic vortex mode is then obtained by integrating
the modal coefficient along the radial direction:
\begin{equation}
C_l
=
\int_{0}^{R_0}
\left| b_l(r,z) \right|^2 r\,dr ,
\end{equation}
where $\left| b_l(r,z) \right|^2$ represents the mode intensities at the radius
$r$ and $R_0$ denotes the integral radius. Accordingly, the mode purity of the
acoustic orbital angular momentum, $P_{\mathrm{OAM}}$, can be calculated as
\begin{equation}
P_{\mathrm{OAM}}
=
\frac{C_l}{\sum_l C_l}.
\end{equation}

As shown in Figs.~S8 and S9, we calculated the mode purities of the acoustic
vortices generated by the $\mathrm{H}_1$ point on different structures. As shown in
Fig.~S8, defined as the energy fraction at the predicted OAM, the mode purity of the acoustic vortex generated by $\mathrm{H}_1$ in
Fig.~4 on the surface of the spherical shell is about 71\%. In addition, for the
ellipsoidal shell under the initial design shown in Fig.~5 of the main text,
the dominant vortex mode with $l = 1$ has a mode purity of approximately 73\%,
as shown in Fig.~S9.

\begin{figure}[htb]
\renewcommand{\figurename}{FIG.}
\renewcommand{\thefigure}{S\arabic{figure}}
\centering
\includegraphics[width=0.6\linewidth]{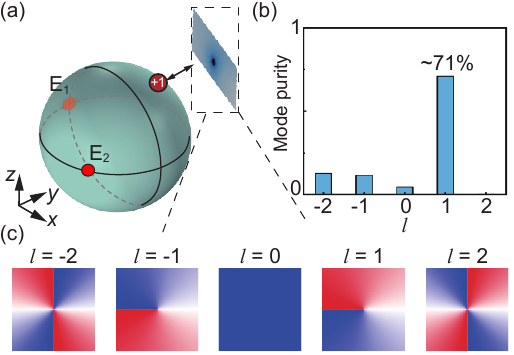}
\caption{\textbf{Mode purity of the acoustic vortex generated by the spherical shell.}
(a) Schematic of a characterized acoustic vortex.
(b) Mode purity of the acoustic vortex shown in (a).
(c) Phase distribution of different orders of OAM modes.
$\mathrm{E}_1$ and $\mathrm{E}_2$ are the point sources.
The parameters are the same as those in Fig.~1 of the main text.}
\label{fig:S8}
\end{figure}

\begin{figure}[htb]
\renewcommand{\figurename}{FIG.}
\renewcommand{\thefigure}{S\arabic{figure}}
\centering
\includegraphics[width=0.6\linewidth]{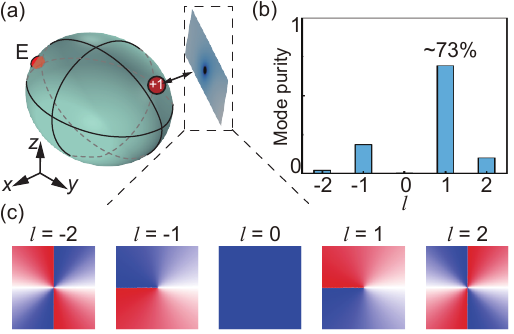}
\caption{\textbf{Mode purity of the acoustic vortex generated by the ellipsoidal shell.}
(a) Schematic of a characterized acoustic vortex.
(b) Mode purity of the acoustic vortex shown in (a).
(c) Phase distribution of different orders of OAM modes.
$\mathrm{E}$ is the point source.
The parameters are the same as those in Fig.~5 of the main text.}
\label{fig:S9}
\end{figure}

\begin{figure}[htb]
\renewcommand{\figurename}{FIG.}
\renewcommand{\thefigure}{S\arabic{figure}}
\centering
\includegraphics[width=0.6\linewidth]{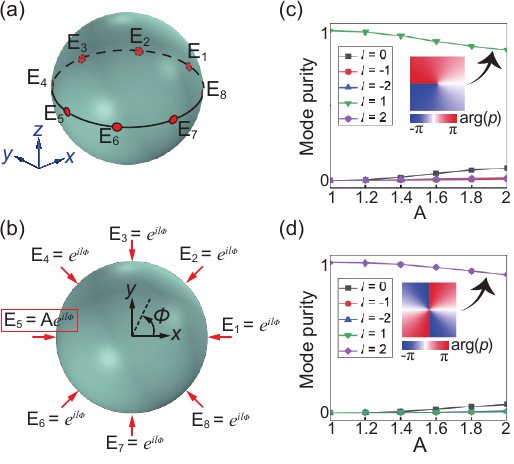}
\caption{\textbf{Improving the acoustic vortex mode purity.}
(a) Schematic of the system to generate an acoustic vortex with eight point sources.
(b) Setting of the point sources.
(c) Dependence of the OAM $l = 1$ mode purity on the amplitude $A$ of source $\mathrm{E}_5$.
(d) Dependence of the OAM $l = 2$ mode purity on $A$.}
\label{fig:S10}
\end{figure}

The mode purity can be improved to $\sim 100\%$ by employing sources with a symmetry, as shown in Fig.~S10, where eight point sources are uniformly distributed on the equator of the spherical shell with the prescribed phase $e^{il\phi}$, where $l \in \mathbb{Z}$ and $\phi$ is the azimuthal coordinate. In Fig.~S10(c), the mode purity of the $l = 1$ mode remains very high even under small perturbations. Here, the perturbation is introduced to the amplitude $A$ of the point source $\mathrm{E}_5$ shown in Fig.~S10(b). Similar robustness can also be observed in Fig.~S10(d) for the generated acoustic vortex with $l = 2$. These results confirm that the singularity can generate highly selective and robust acoustic OAM states.

\vspace{1em}

\noindent\textbf{2.~Radiation efficiency}

We also evaluated the acoustic radiation efficiency numerically. The efficiency is defined as
\begin{equation}
\eta =
\frac{P^R}{P^R + \mathrm{Abs}},
\end{equation}
where $P^R$ is the acoustic power radiated into free space and can be obtained by integrating the power flux density over a closed surface enclosing the spherical shell, as illustrated in Fig.~S11(a). The term $\mathrm{Abs}$ accounts for the absorption of elastic wave power within the shell. Figure S11(b) presents the power spectra of acoustic radiation and elastic absorption, which exhibit three distinct peaks corresponding to the system's eigenmodes. The radiation efficiency is shown in Fig.~S11(c). At 2100~Hz, the working frequency in the main text, the efficiency is around 5\%. The radiation efficiency can be enhanced near the eigenmode resonances, reaching about 15\%. This indicates that, in practical applications, resonant excitation can be exploited to enhance the efficiency of acoustic vortex generation. It should be noted that the present design was not intentionally optimized for radiation efficiency; rather, the primary focus was on demonstrating acoustic vortex generation. 

\begin{figure}[htb]
\renewcommand{\figurename}{FIG.}
\renewcommand{\thefigure}{S\arabic{figure}}
\centering
\includegraphics[width=\linewidth]{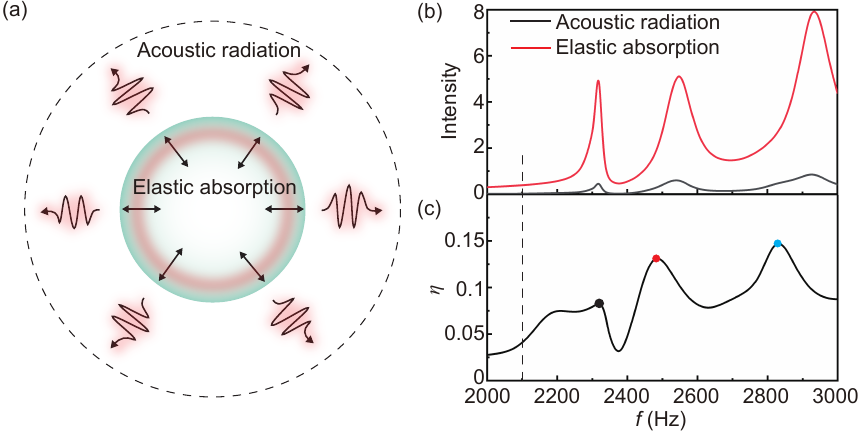}
\caption{(a) Schematic of the acoustic radiation and elastic absorption.
(b) Acoustic radiation spectrum and elastic absorption spectrum.
The black-dashed line marks the working frequency $f = 2100~\mathrm{Hz}$.
(c) Spectrum of acoustic radiation efficiency.
The spherical shell has a radius of $R = 8~\mathrm{cm}$ and a thickness of $t = 0.3~\mathrm{cm}$.}
\label{fig:S11}
\end{figure}

\vspace{1em}

\noindent\textbf{3.~Bandwidth and robustness}

For the spherical-shell system driven by two out-of-phase source excitations, the hybrid singularities $\mathrm{H}_1$, corresponding to the flexural wave singularities that induce the acoustic vortices in free space, remain stable on the shell surface in the frequency range of 1300-2500 Hz (corresponding to the fractional bandwidth 63\%). Figure S12 shows the flexural wave singularities and the dislocation lines of the pressure field at three representative frequencies: 1300 Hz, 1900 Hz, and 2500 Hz. We observe that varying the excitation frequency causes a spatial shift of the singularities, while the acoustic vortex generation is preserved. When the entire surface of the spherical shell is considered, $\mathrm{H}_1$ singularities and the associated acoustic vortex generation can occur at all investigated frequencies, owing to the persistent out-of-phase modal interference introduced by the sources. This broadband robustness is governed by the underlying physics of modal interference.

\begin{figure}[htb]
\renewcommand{\figurename}{FIG.}
\renewcommand{\thefigure}{S\arabic{figure}}
\centering
\includegraphics[width=0.7\linewidth]{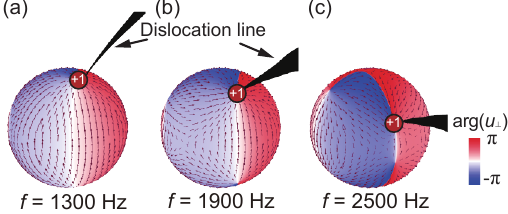}
\caption{\textbf{Acoustic vortex generation on the spherical shell at different frequencies.}
(a) $f = 1300~\mathrm{Hz}$, 
(b) $f = 1900~\mathrm{Hz}$, and 
(c) $f = 2500~\mathrm{Hz}$.
The red arrows denote the spin.
The spherical shell has the same geometric parameters as those in Fig.~1 of the main text.}
\label{fig:S12}
\end{figure}
\begin{figure}[htb]
\renewcommand{\figurename}{FIG.}
\renewcommand{\thefigure}{S\arabic{figure}}
\centering
\includegraphics[width=0.7\linewidth]{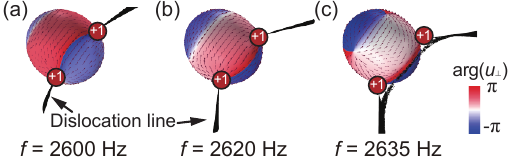}
\caption{\textbf{Acoustic vortex generation on the ellipsoidal shell at different frequencies.}
(a) $f = 2600~\mathrm{Hz}$, 
(b) $f = 2620~\mathrm{Hz}$, and 
(c) $f = 2635~\mathrm{Hz}$.
The red arrows denote the spin.
The ellipsoidal shell has the same geometric parameters as those in Fig.~5 of the main text.}
\label{fig:S13}
\end{figure}

For the ellipsoidal shell excited by a single source, $\mathrm{H}_1$ exists within the frequency range of 2600--2635~Hz. Figure~S13 shows the hybrid singularities $\mathrm{H}_1$, together with the dislocation lines of the
acoustic pressure field, at three representative frequencies: 2600~Hz,
2620~Hz, and 2635~Hz. The background color represents the phase of the
flexural wave, $\arg(u_{\perp})$. The emergence of hybrid singularities in
this case is attributed to interference between two excited eigenmodes with a
$\pi/2$ phase difference. At $f = 2635$~Hz, the singularities and the
dislocation lines of the pressure field begin to merge in the near field, and
the acoustic vortices disappear at higher frequencies.

Conventional approaches to acoustic vortex generation, such as spiral phase
plates and phased arrays, rely on phase engineering to achieve vortex phase
winding, which entails complex structural designs and is inherently vulnerable
to perturbations. In contrast, the present mechanism can be implemented in
simple structures, e.g., spheres or ellipsoids, and arises fundamentally from
elastic-wave topological singularities, making it robust to small perturbations.

\vspace{1em}

\noindent\textbf{4.~Acoustic vortices inside the shell}

In the main text, we have focused on the acoustic vortices generated outside the shells in free space. Such vortices also emerge inside the shells. Figure S14(a) shows the flexural field singularities on the inner surface of the spherical shell, under the same excitation as in Fig. 1 of the main text. These singularities also imprint dislocation lines on the pressure field within the shell, thereby generating acoustic vortices, as shown in Fig. S14(b). Figure S14(c) shows the flexural field singularities on the inner surface of the ellipsoidal shell, under the same excitation as in Fig. 5 of the main text. Similarly, these singularities generate acoustic vortices inside the shell, as shown in Fig. S14(d). The generation of acoustic vortices within the shells is attributed to the coupling between elastic and acoustic waves, akin to the mechanism that underlies their generation outside the shells.

\begin{figure}[htb]
\renewcommand{\figurename}{FIG.}
\renewcommand{\thefigure}{S\arabic{figure}}
\centering
\includegraphics[width=0.7\linewidth]{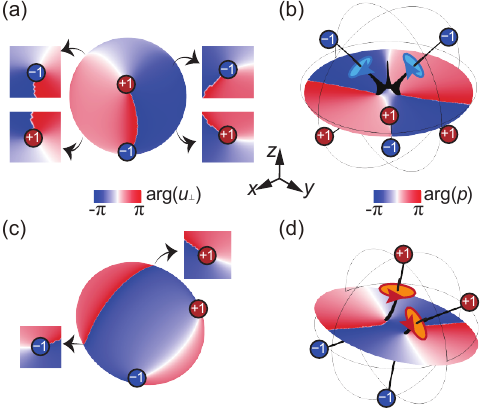}
\caption{\textbf{Acoustic vortices inside the shell.}
(a) Flexural field singularities on the inner surface of the spherical shell.
(b) Acoustic vortices and dislocation lines inside the spherical shell.
(c) Flexural field singularities on the inner surface of the ellipsoidal shell.
(d) Acoustic vortices and dislocation lines inside the ellipsoidal shell.
The excitations in (a, b) and (c, d) are the same as those in Fig.~1 and Fig.~5 of the main text, respectively.}
\label{fig:S14}
\end{figure}

\vspace{1em}

\noindent\textbf{5.~Further experimental measurement of the acoustic vortex }

Besides the acoustic vortex generated on the surface of the ellipsoidal shell.
Further experiments have been performed to verify the acoustic vortex with
topological charge $l = -1$ generated via another $\mathrm{H}_1$ point, and the
results are summarized in Fig.~S15. The comparison between numerical and
experimental results is summarized in Fig.~S15. Both simulations and
measurements confirm the presence of a vanishing flexural amplitude and a phase
singularity, as shown in Figs.~S15(c) and S15(d). This $\mathrm{H}_1$ point
generates an acoustic vortex in free space, as shown in Fig.~S15(e), where the
simulated pressure field exhibits a clear singularity and phase winding pattern.
Figure~S15(f) shows the simulation, solid lines, and experimental, dots, results
for the pressure amplitude and phase along the circular array. We clearly
observe the $-2\pi$ phase winding corresponding to a topological charge
$l = -1$, with excellent agreement between the simulation and experimental
results. These results provide further evidence for the radiated acoustic
vortex, with direct experimental evidence.

\begin{figure}[htb]
\renewcommand{\figurename}{FIG.}
\renewcommand{\thefigure}{S\arabic{figure}}
\centering
\includegraphics[width=0.7\linewidth]{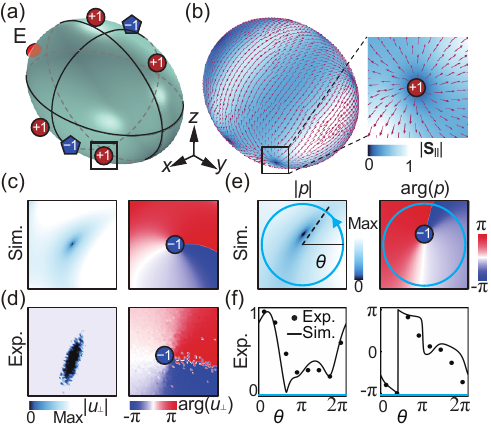}
\caption{\textbf{Acoustic vortex generation by an ellipsoidal shell driven by a point source.}
(a) Hybrid singularities on the shell surface.
(b) Distribution of in-plane spin.
Numerically simulated (c) and experimentally measured (d) intensity and phase distribution of the flexural wave near the $\mathrm{H}_1$ point marked in (a).
(e) Simulated pressure field in the measured plane.
(f) Experimentally measured pressure field along the blue circle in (e).}
\label{fig:S15}
\end{figure}

\end{document}